\documentclass[aps,preprintnumbers,amsmath,amssymb,nofootinbib,superscriptaddress,notitlepage,11pt]{revtex4-1}
\usepackage[utf8]{inputenc}
\usepackage{comment}
\usepackage{amssymb}
\usepackage{color}
\usepackage{rotating}

\usepackage{amsmath}
\usepackage{ulem}
\usepackage{overpic}
\usepackage{graphicx}
\usepackage{dcolumn}
\usepackage{graphicx}
\usepackage{bm}
\usepackage{chngpage}
\usepackage{multirow}
\usepackage{slashed}
\usepackage{indentfirst}
\usepackage{booktabs}
\usepackage{amssymb,amsfonts}
\usepackage{amsmath}
\usepackage{color}
\usepackage{hyperref}
\usepackage{cleveref}

\begin{document}

\title{Hidden charm pentaquark with strangeness $P_{cs}^*(4739)$ as a $\Sigma_{c}\bar{D}\bar{K}$ bound state}

\author{Tian-Wei Wu}

\affiliation{School of Physics,
  Beihang University, Beijing 100191, China}

  \author{Ya-Wen Pan}
  \affiliation{School of Physics,
  Beihang University, Beijing 100191, China}
\author{Ming-Zhu Liu}
\affiliation{School of Space and Environment, Beihang University, Beijing 102206, China}
\affiliation{School of Physics,
  Beihang University, Beijing 100191, China}
\author{Jun-Xu Lu}
\affiliation{School of Space and Environment, Beihang University, Beijing 102206, China}
\affiliation{School of Physics,
  Beihang University, Beijing 100191, China}

\author{Li-Sheng Geng}
\email{Corresponding author: lisheng.geng@buaa.edu.cn}
\affiliation{School of Physics,
  Beihang University, Beijing 100191, China}
\affiliation{
Beijing Key Laboratory of Advanced Nuclear Materials and Physics,
Beihang University, Beijing 102206, China}
\affiliation{School of Physics and Microelectronics, Zhengzhou University, Zhengzhou, Henan 450001, China}

\author{Xiao-Hai Liu}
\email{Corresponding author: xiaohai.liu@tju.edu.cn}
\affiliation{Center for Joint Quantum Studies and Department of Physics, School of Science, Tianjin University, Tianjin 300350, China}

\date{\today}


\begin{abstract}
Motivated by the recent discovery of the first hidden charm pentaquark state with strangeness $P_{cs}(4459)$ by the LHCb Collaboration, we study the likely existence of a three-body $\Sigma_{c}\bar{D}\bar{K}$ bound state, which shares the same minimal quark content as $P_{cs}(4459)$.  The $\Sigma_{c}\bar{D}$ and $DK$ interactions are determined by reproducing $P_c(4312)$ and $D_{s0}^*(2317)$ as $\Sigma_c\bar{D}$ and $\bar{D}\bar{K}$ molecules, respectively, while the $\Sigma_c\bar{K}$ interaction is constrained by chiral effective theory. We indeed find a three-body bound state by solving the  Schr\"odinger equation using the Gaussian Expansion Method, which can be viewed as an excited  hidden charm pentaquark state with strangeness, $P_{cs}^*(4739)$, with $I(J^P)=1(1/2^+)$ and  a binding energy of $77.8^{+25}_{-10.3}$ MeV. We further study its strong decays via triangle diagrams  and show that its partial decay widths into $D\Xi_c'$ and $D_s^*\Sigma_c$ are of a few ten's MeV, with the former being dominant.

\end{abstract}

\maketitle

\section{Introduction}

Since 2003, a large number of exotic hadronic states have been discovered, which attracted a lot of attention both theoretically and experimentally. Though their nature is largely undetermined, the hadronic molecular interpretation of these states has become rather popular because of the fact that many of them are located close to the thresholds of two conventional hadrons.  For example, the well-known $X(3872)$,  which is just located at the $\bar{D} D^*$ threshold, can be naturally explained as
a $D^*\bar{D}$ molecule~\cite{Swanson:2003tb}. This molecular picture can also well explain its isospin-breaking decay~\cite{Abe:2005ix,delAmoSanchez:2010jr,Li:2012cs}, but is disfavored by its radiative decay to $J/\psi \gamma$~\cite{Aubert:2008ae,Bhardwaj:2011dj,Li:2009zu}. Meanwhile, there are some  claims that $X(3872)$ is not a pure molecule~\cite{Coito:2012vf}, but a hybrid state of charmonium and hadronic molecule~\cite{Badalian:2012jz,Takizawa:2012hy}. It is interesting to note that after about 20 years' study, the nature of $X(3872)$ still remains unclear and  more clarifications are needed both theoretically and experimentally.

Another intriguing exotic state is  $D_{s0}^*(2317)$~\cite{Aubert:2003fg,Besson:2003cp,Krokovny:2003zq}, which is located  45 MeV below the $DK$ threshold and has a decay width less than 3.8 MeV. The observed mass and width are far away from the predicted mass and width  in the naive quark model, which are about 2460 MeV and hundreds MeV respectively~\cite{Barnes:2003dj}, thus  $D_{s0}^*(2317)$ is difficult to be interpreted as a conventional $c\bar{s}$ meson. On the other hand, due to the strongly attractive $DK$ interaction predicted by  chiral perturbation theory and lattice QCD~\cite{Altenbuchinger:2013vwa,Mohler:2013rwa},  $D_{s0}^*(2317)$ can  be easily explained as a $DK$ molecule~\cite{Kolomeitsev:2003ac,Hofmann:2003je,Guo:2008gp,Guo:2006fu,Guo:2009ct,MartinezTorres:2011pr,Yao:2015qia,Guo:2015dha,Guo:2018kno,Albaladejo:2018mhb,Altenbuchinger:2013vwa,Geng:2010vw,Wang:2012bu,Liu:2009uz,Guo:2018tjx,Liu:2012zya,Mohler:2013rwa,Lang:2014yfa,Bali:2017pdv}. The same molecular picture is adopted to study the pentaquark states, i.e.,  $P_c(4312)$, $P_c(4440)$, and $P_c(4457)$, discovered by the LHCb Collaboration~\cite{Aaij:2019vzc}. These pentaquark states are explained as $\Sigma_c\bar{D}^{(*)}$ molecules due to the fact they are close to the thresholds of $\Sigma_c\bar{D}^{(*)}$~\cite{Liu:2019tjn,Chen:2019asm,He:2019ify,Chen:2019bip,Xiao:2019aya,Guo:2019kdc,Xiao:2019mst,Guo:2019fdo,Eides:2019tgv,Cheng:2019obk,Wang:2019got,Meng:2019ilv,Weng:2019ynv,Du:2019pij,Burns:2019iih,Liu:2019zvb}.

The molecular explanation of some exotic states can be extended by symmetries, such as the heavy quark symmetry and the $SU(3)$ flavor symmetry. The $D_{s1}(2460)$ state, the heavy quark spin partner of $D_{s0}^*(2317)$, can also be interpreted as a $D^*K$ molecule with heavy quark spin symmetry~\cite{Geng:2010vw,Lang:2014yfa}. In this doublet molecular picture the mass splitting of $D_{s0}^*(2317)$ and $D_{s1}(2460)$ can be easily understood, which supports the molecular interpretation of $D_{s0}^*(2317)$ and $D_{s1}(2460)$.   Recently, a new structure is observed in the $J/\psi\Lambda$ invariant mass distribution of the $\Xi_{b}^-\rightarrow J/\psi\Lambda K^-$ decay by the LHCb Collaboration~\cite{Aaij:2020gdg}. This  structure is consistent with a charmonium pentaquark state with strangeness denoted as $P_{cs}(4459)$, which could be viewed as a SU(3)-flavor symmetry partner of the $P_c$ pentaquark states~\cite{Peng:2020hql,Liu:2020hcv,Xiao:2021rgp,Wu:2021dmq,Lu:2021irg}.
Motivated by this new observation and the molecular picture for $D_{s0}^*(2317)$ and the $P_c$ states, we study the $\Sigma_{c}\bar{D}\bar{K}$ three-body system, whose minimum quark content is $c\bar{c}sqq$, the same as  $P_{cs}(4459)$, to check whether there exist hidden charm fermionic three-body bound states, and to explore its possible decays.

It should be mentioned that hadronic few-body systems have been extensively studied by different methods in the charm sector, such as the  hidden charmed $D\bar{D}K$~\cite{Wu:2020job} and $D\bar{D}^*K$~ \cite{Wu:2020job,Ma:2017ery,Ren:2018pcd}, the $BB\bar{K}$ system~\cite{Wu:2021ljz}, the $\pi D\bar{D}$~\cite{Baru:2011rs}, $\rho D\bar{D}$~\cite{Durkaya:2015wra}, $BD\bar{D}$~\cite{Dias:2017miz} and $\Xi_{cc}\bar{\Xi}_{cc}\bar{K}$~\cite{Wu:2020rdg}, the singly charmed $DNN$~\cite{Bayar:2012dd}, $DK\bar{K}$~\cite{MartinezTorres:2012jr,Debastiani:2017vhv}, and $NDK$($ND\bar{K}$)~\cite{Xiao:2011rc}, the doubly charmed $DDK$~\cite{MartinezTorres:2018zbl,Wu:2019vsy,Huang:2019qmw,Pang:2020pkl}, $BDD$~\cite{Dias:2017miz} and $DD^*K$~\cite{Ma:2017ery} systems, the triply charmed four-body $DDDK$~\cite{Wu:2019vsy}, and the quadruply charmed $\Xi_{cc}\Xi_{cc}\bar{K}$~\cite{Wu:2020rdg} systems. For recent reviews, see Refs.~\cite{MartinezTorres:2020hus,Wu:2021dwy}.

The decay of three-body bound states have also attracted much attention. In Ref.~\cite{Huang:2019qmw}, the decay of the $DDK$ bound state has been studied  via the triangle mechanism.
From the conclusion of recent works~\cite{Wu:2019vsy,MartinezTorres:2018zbl}, the $DDK$ bound state is mainly made of $DD_{s0}^*(2317)$, accordingly the decay width of $DDK$ can be estimated through $DD_{s0}^*(2317)$ to $DD_{s}^{\ast}$/$D^{\ast}D_{s}$ by exchanging $K$/$\eta$~\cite{Huang:2019qmw}.  Using the same approach  the decay width of $D\bar{D}^{(*)}K$ was also calculated~\cite{Ren:2019umd,Wu:2020rdg}. In addition, the $D\bar{D}^{*}K$ production rate in the $B$ meson decay was  studied in Ref.~\cite{Ren:2019rts}.
Although there are many studies on few-body systems in the heavy hadron sector, the  $\Sigma_{c}\bar{D}\bar{K}$ system could generate the first hidden charm fermionic three-body bound state, which is likely to be found at the current facilities, especially considering the successful discoveries of the $P_c$ and $P_{cs}$ states.

The manuscript is organized as follows, In Sec.~\ref{Sec:Interactions}, we explain how we parametrize and determine the two-body interaction inputs. In Sec.~\ref{Sec:GEM}, we explain how to construct the three-body wave functions and solve the $\Sigma_{c}\bar{D}\bar{K}$ three-body system.
In Sec.~\ref{Results}, we present our predictions of the $\Sigma_{c}\bar{D}\bar{K}$ bound state and study its strong decay. Finally, a short summary is given in Sec.~\ref{Summary}.

\section{Two-body Interactions}
\label{Sec:Interactions}
The study of a three-body system depends on the sub two-body interactions. For the $\Sigma_{c}\bar{D}\bar{K}$ system, we need to know the $\Sigma_{c}\bar{D}$, $\bar{D}\bar{K}$, and $\Sigma_{c}\bar{K}$ interactions. The $\Sigma_{c}\bar{D}$ and $DK$ interactions, as mentioned in the Introduction, are attractive enough to form bound states, namely $P_c(4312)$ and $D_{s0}^*(2317)$. Therefore we could determine the interactions by reproducing the  two states. For the case of the $\Sigma_{c}\bar{K}$ system, there is no  such information, we would resort to  chiral perturbation theory and relate it to the $\bar{K}N$ interaction. For the masses of the particles used in the present study, we refer to Table~\ref{Masses}.

\begin{table}[htpb]
\centering
\caption{ Hadron masses needed in this work   (in units of GeV).\label{Masses}}\label{table0}
\begin{tabular}{ccccccc}
\hline\hline
   Meson    &~~~~~  $D_{s0}^{-}$        &~~~~~  $D_{s}^{\ast-}$                   &~~~~~  $K^{-}$                    &~~~~~ $\bar{K}^{0}$       &~~~~~ $D^{-}$          &~~~~~$\bar{D}^{0}$ \\
Mass  &~~~~~ $2.3178$ &~~~~~ $2.1122$                &~~~~~ $0.493677$                   &~~~~~ $0.497611$       &~~~~~ $1.86965$        &~~~~~$1.86483$   \\  \hline
Baryon  &~~~~~  $\Xi_{c}^{\prime+}$        &~~~~~  $\Xi_{c}^{\prime0}$                    &~~~~~  $\Sigma_{c}^{++}$                    &~~~~~ $\Sigma_{c}^{+}$       &~~~~~ $\Sigma_{c}^{0}$         &~~~~~$P_{c}$ \\
Mass  &~~~~~ $2.5774$ &~~~~~ $2.5788$                &~~~~~ $2.45397$                   &~~~~~ $2.4529$       &~~~~~ $2.45375$        &~~~~~$4.3119$   \\ \hline
                     \hline
\end{tabular}
\end{table}

For the $\Sigma_{c}\bar{D}$ interaction, we  refer to the contact-range effective field theory of Ref.~\cite{Liu:2019tjn}, in which the $\Sigma_{c}^{(*)}\bar{D}^{(*)}$ interactions are  constructed to explain $P_c(4312)$, $P_c(4440)$ and $P_c(4457)$ as part of a heavy quark spin symmetry (HQSS) multiplet. The $\Sigma_{c}\bar{D}$ potential in the contact-range effective field theory reads
\begin{eqnarray}
\label{Poten:SigmacDbar}
V(\frac{1}{2}^-,\Sigma_{c}\bar{D})=C_a,
\end{eqnarray}
with $C_a$ a coupling constant. In Ref.~\cite{Liu:2019tjn}, the authors regularized the potential with a separable form factor and a cutoff $\Lambda$ in momentum space and allowed the couplings to depend on the cutoff
\begin{eqnarray}
\langle p|V_{\Lambda}|p'\rangle=C_{\Lambda}f\left(\frac{p}{\Lambda}\right)f\left(\frac{p'}{\Lambda}\right).
\end{eqnarray}
Here we propose a similar treatment in coordinate space. The contact potential of Eq.~(\ref{Poten:SigmacDbar}) in coordinate space can be obtained by Fourier transformation
\begin{eqnarray}
  V_{\Sigma_{c}\bar{D}}(\vec{r}) = C_a\delta^{(3)}(\vec{r}) \, ,
  \label{eq:V-SD}
\end{eqnarray}
which is singular and requires regularization.
For this purpose we choose a Gaussian regulator of the type
\begin{eqnarray}
  V_{\Sigma_{c}\bar{D}}(\vec{r}) = C_a\frac{e^{-(r/R_a)^2}}{\pi^{3/2} R_a^3} \, ,
\end{eqnarray}
where $R_c$ is the cutoff we use to smear the delta function.
However the previous expression is still problematic, as the prediction  of a $\Sigma_{c}\bar{D}$ bound state and its binding energy depends on both the coupling $C_a$ and the cutoff. This can
be solved by taking $C_a$ cutoff dependent and therefore one is left with a renormalized potential
\begin{eqnarray}
V_{\Sigma_{c}\bar{D}}({r}; R_a) = C(R_a)e^{-(r/R_a)^2}
\end{eqnarray}
with $R_a$ the cutoff and $C(R_a)$ the running coupling constant fixed by fitting to the bound state of $\Sigma_{c}\bar{D}$ with a  binding energy of 8.9 MeV, which corresponds to $P_c(4312)$.

The most important contribution to the $DK$ interaction
is the Weinberg-Tomozawa term between a kaon and a  charmed meson~\cite{Altenbuchinger:2013vwa}, which in  non-relativistic normalization reads
\begin{eqnarray}
  V_{WT}(DK) = - \frac{C_{WT}(I)}{2 f_{\pi}^2} \, ,
\end{eqnarray}
with $f_{\pi} \simeq 130\,{\rm MeV}$ and $C_{WT}(0) = 2$, $C_{WT}(1) = 0$
for the isoscalar and isovector channels, respectively.
Following the same logic in treating  the $\Sigma_{c}\bar{D}$ interaction, the isoscalar contact-range $DK$ interaction can be Fourier transferred into coordinate space and represented by a Gaussian shape potential that was already adopted in our previous works~\cite{Wu:2019vsy,Wu:2020job}
\begin{eqnarray}
\label{Poten:DK}
V_{DK}({r}; R_b) = C(R_b)e^{-(r/R_b)^2},
\end{eqnarray}
where $R_b$ is a cutoff, $C(R_b)$ is a running coupling constant related to $R_b$, which can be determined by fitting to a binding energy of $45$ MeV for the $DK$ bound state corresponding to $D_{s0}^*(2317)$.

For the $\Sigma_{c}\bar{K}$ interaction, we will resort to the unitarized
chiral perturbation theory developed in Ref.~\cite{Lu:2014ina} to describe the interactions between a ground-state singly charmed (bottom) baryon and a pseudo-Nambu–Goldstone boson. The leading order chiral effective Lagrangian reads
\begin{eqnarray}
\label{Lag:SigmacKbar}
\mathcal{L} = \frac{i}{16f_0^{2}}\mathrm{Tr}(\bar{H}_{[6]}(x)\gamma^{\mu}[H_{[6]}(x),[\phi(x),(\partial_{\mu}\phi(x))]_{-}]_{+}),
\end{eqnarray}
where the $\bar{H}_{[6]}$ and $\phi$ collect the  sextet charmed baryons and Goldstone bosons respectively(for details we refer to Ref.~\cite{Lu:2014ina}). The Lagrangian above leads to the well-known Weinberg-Tomozawa term
\begin{equation}\label{VSigK:WT}
V_{WT}(\Sigma_{c}\bar{K})=\frac{C_{ij}}{4f_0^2}(\slashed{k}_2+\slashed{k}_4)
\end{equation}
with the coupling $C_{ij}=-3$, $k_2$ and $k_4$ the momentum of the incoming and outgoing Kaons. Neglecting  subleading corrections~\footnote{In Ref.~\cite{Wu:2019vsy}, we constructed a repulsive core to describe the NLO repulsive contribution to the $DK$ interaction, and found that the subleading correction only influences the three-body binding energies by less than 1 MeV, and thus can be neglected.}, the $\Sigma_{c}\bar{K}$ interaction is the same as the $N\bar{K}$ interaction (see, for example, those in Refs.~\cite{Oller:2000fj,Oller:2006jw,Borasoy:2005ie}). Thus the $\Sigma_{c}\bar{K}$ potential is taken to be of the same form as the $N\bar{K}$ potential of Ref.~\cite{Wu:2020rdg}
\begin{equation}
V_{\Sigma_{c}\bar{K}}(r;R_c)\simeq C(R_c)e^{-(r/R_c)^2},
\end{equation}
while the coupling $C(R_c)$ is determined by reproducing the binding energy of $\Lambda(1405)$ as a $N\bar{K}$ bound state.
To estimate the uncertainties of the  $\Sigma_{c}\bar{D}$, $DK$, and $\Sigma_{c}\bar{K}$ interactions, we vary the cutoffs $R_a$, $R_b$, and $R_c$ from 0.5 to 2.0 fm. In principle, the cutoffs can be different for each sub two-body system. However, because the uncertainties of this system  mainly originate from the cutoff $R_c$ (see  Sec.~\ref{Results}), we choose the cutoffs $R_a$ and $R_b$ the same as $R_c$ ranging from 0.5 to 2.0 fm in the following numerical study.

\section{Gaussian Expansion Method}
\label{Sec:GEM}

\begin{figure}[!h]
		\centering
		
	\begin{overpic}[scale=.5]{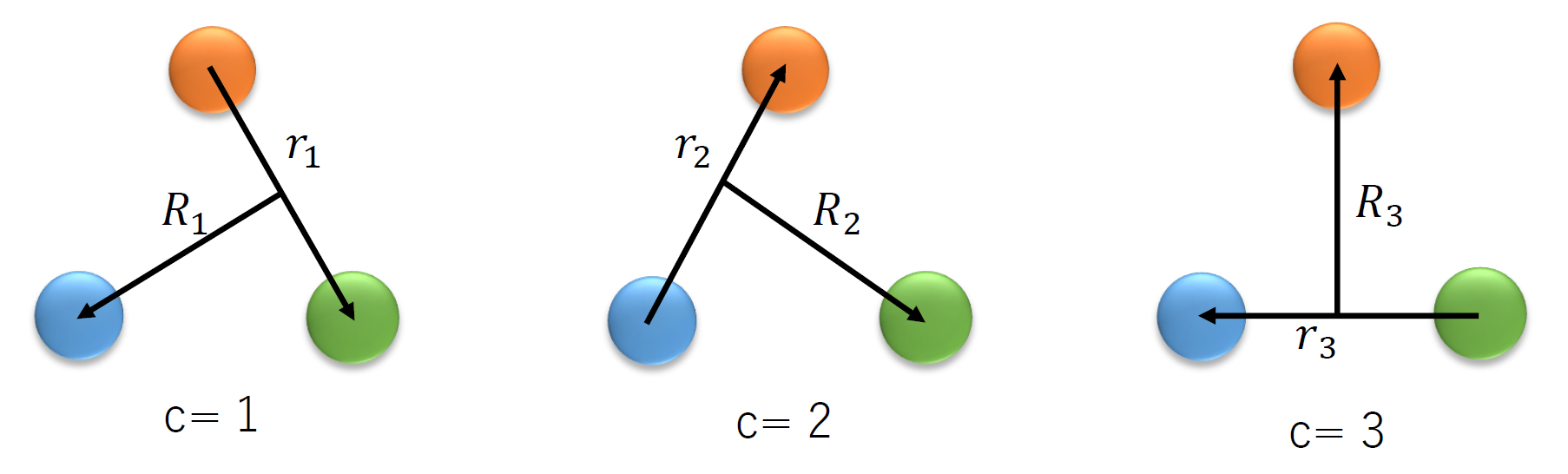}
		\put(-1,9){$\Sigma_{c}$}	\put(26,9){$\bar{K}$}  \put(13,29){$\bar{D}$}
		\put(35,9){$\Sigma_{c}$}	\put(62,9){$\bar{K}$}  \put(49,29){$\bar{D}$}
		\put(70.5,9){$\Sigma_{c}$}	\put(98,9){$\bar{K}$}  \put(84.5,29){$\bar{D}$}
		\end{overpic}

		\caption{Three permutations of the Jacobi coordinates for the $\Sigma_{c}\bar{D}\bar{K}$ system.}\label{Jac}
	\end{figure}
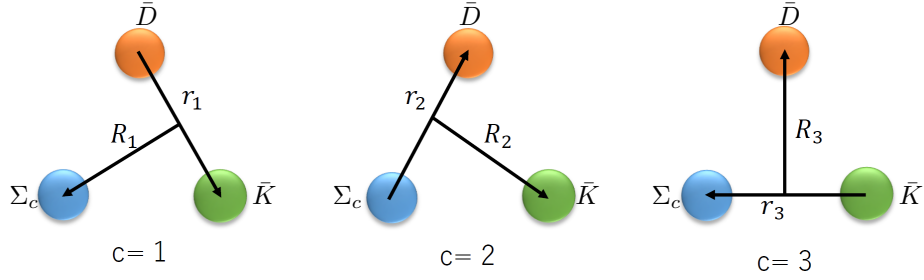

Once all the relevant sub two-body interactions are fixed as specified above, we employ the Gaussian Expansion Method (GEM) \cite{Hiyama:2003cu} to solve the Schr\"{o}dinger equation to study the three-body $\Sigma_{c}\bar{D}\bar{K}$ system. The corresponding Schr\"{o}dinger equation  is
\begin{equation}\label{schd}
\hat{H}\Psi_{JM}^{total}=E\Psi_{JM}^{total},
\end{equation}
with the following Hamiltonian
\begin{equation}\label{hami}
\hat{H}=\sum_{i=1}^{3}\frac{p_i^2}{2m_i}-T_{c.m.}+V_{\bar{D}\bar{K}}(r_1)+V_{\Sigma_{c}\bar{D}}(r_2)+V_{\bar{K}\Sigma_{c}}(r_3),
\end{equation}
where $T_{c.m.}$ is the kinetic energy of the center of mass and $V(r)$'s are the potentials between the two relevant particles.
The three Jacobian coordinates for the $\Sigma_{c}\bar{D}\bar{K}$ system are shown in Fig.~\ref{Jac}.
	The total wave function  is  a sum of the amplitudes of the three rearrangement channels ($c = 1-3$) written in Jacobian coordinates,
	\begin{equation}\label{Sch}
	\Psi_{JM}^{total}=\sum_{c,\alpha}C_{c,\alpha}\Psi_{JM,\alpha}^{c}(\mathbf{r}_c,\mathbf{R}_c)
	\end{equation}
	where $\alpha=\{nl,NL,\Lambda,tT\}$ and $C_{c,\alpha}$ are the expansion coefficients. Here $l$ and $L$ are the orbital angular momentum for the coordinates $r$ and $R$,  $t$ is the isospin of the  two-body subsystem in each channel, $\Lambda$ and $T$ are the total orbital angular momentum and isospin, $n$ and $N$ are the numbers of Gaussian basis functions corresponding to coordinates $r$ and $R$, respectively.
	The wave function of  each channel has the following form
	\begin{equation}\label{dd}
	\Psi_{JM,\alpha}^{c}(\mathbf{r}_c,\mathbf{R}_c)=H_{T,t}^c\otimes[\Phi_{lL,\Lambda}^c]_{JM},
	\end{equation}
	where $H_{T,t}^c$ is the isospin wave function and $\Phi_{lL,\Lambda}^c$ the orbital wave function. The total isospin wave function in each channel reads
	\begin{equation}\label{isospinwave}
	\begin{split}
	H_{T,t}^{c=1}& =[[\eta_{\frac{1}{2}}(\bar{D})\eta_{\frac{1}{2}}(\bar{K})]_{t_1}\eta_{1}(\Sigma_{c})]_{T}, \\
	H_{T,t}^{c=2}& =[[\eta_{1}(\Sigma_{c})\eta_{\frac{1}{2}}(\bar{D})]_{t_2}\eta_{\frac{1}{2}}(\bar{K})]_{T},\\
	H_{T,t}^{c=3}& =[[\eta_{\frac{1}{2}}(\bar{K})\eta_{1}(\Sigma_{c})]_{t_3}\eta_{\frac{1}{2}}(\bar{D})]_{T},
	\end{split}
	\end{equation}
where $\eta$ is the isospin wave function of each particle.
The isospin factors of this system are listed in Table~\ref{ISF}.
		\begin{table}{]}
		\caption{Isospin factors used in calculating the Hamiltonian matrix elements.}
			\label{ISF}
		\begin{tabular}{c|ccc}
			\hline
			\hline
				&$H_{1,0}^{c=1}$&$H_{1,1/2}^{c=2}$&$H_{1,1/2}^{c=3}$\\
			\hline
			$H_{1,0}^{c=1}$&1&$-\frac{1}{\sqrt{3}}$&$-\frac{1}{\sqrt{3}}$\\
			$H_{1,1/2}^{c=2}$&$-\frac{1}{\sqrt{3}}$&1&$-\frac{1}{3}$\\
			$H_{1,1/2}^{c=3}$&$-\frac{1}{\sqrt{3}}$&$-\frac{1}{3}$&1\\
			\hline
			\hline
		\end{tabular}
	\end{table}
	The orbital wave function $\Phi_{lL,\Lambda}^c$ is given in terms of the Gaussian basis functions
	\begin{equation}\label{nj}
	\Phi_{lL,\Lambda}^c(\mathbf{r}_c,\mathbf{R}_c)=[\phi_{n_cl_c}^{G}(\mathbf{r}_c)\psi_{N_cL_c}^{G}(\mathbf{R}_c)]_{\Lambda},
	\end{equation}
	\begin{equation}\label{nj}
	\phi_{nlm}^{G}(\mathbf{r}_c)=N_{nl}r_c^le^{-\nu_n r_c^2} Y_{lm}({\hat{r}}_c),
	\end{equation}
	\begin{equation}\label{nj}
	\psi_{NLM}^{G}(\mathbf{R}_c)=N_{NL}R_c^Le^{-\lambda_n R_c^2} Y_{LM}({\hat{R}}_c).
	\end{equation}
	Here $N_{nl}(N_{NL})$ is the normalization constant of the Gaussian basis and the parameters $\nu_n$ and $\lambda_n$ are given by
	\begin{equation}\label{vn}
	\begin{split}
	\nu_n &=1/r_n^2,\qquad r_n=r_{min}a^{n-1}\quad (n=1,n_{max}), \\
	\lambda_N &=1/R_N^2,\quad R_N=R_{min}A^{N-1}\quad (N=1,N_{max}),
	\end{split}
	\end{equation}
	where $\{n_{max},r_{min},a$ or $r_{max}\}$ and  $\{N_{max},R_{min},A$ or $R_{max}\}$ are gaussian basis parameters.

	\begin{table}[!h]
	\caption{Quantum numbers and the numbers of Gaussian basis used in each Jacobi coordinate channel $(c=1-3)$ of the $\Sigma_{c}\bar{D}\bar{K}$ $I(J^P)=1(\frac{1}{2}^+)$ system. }\label{Conf}
	\centering
	\begin{tabular}{c c c c c c c c c c c}
		\hline\hline
		channels &coupling types& $l$ &$L$& $\Lambda$ & $t$ & $T$ & $J$ & $P$ &$n_{max}$ & $N_{max}$\\
		\hline
		1 & $(\bar{D}\bar{K})\Sigma_{c}$ &0& 0  & 0 & 0 & 1 &1/2& $+$&10&10 \\
		2 & $(\Sigma_{c}\bar{D})\bar{K}$ &0&0 & 0 & 1/2 & 1 & 1/2& $+$&10&10 \\
		3 & $(\bar{K}\Sigma_{c})\bar{D}$ &0 &0 & 0 & 1/2 &1 & 1/2& $+$ &10&10\\
		\hline\hline
	\end{tabular}
\end{table}

	With the constraints of $D_{s0}^*(2317)$ as a $DK$ bound state with quantum numbers $I(J^P)=0(0^+)$ and $\Sigma_{c}$ a $1(1/2^+)$ particle, the quantum numbers of the three-body $\Sigma_{c}\bar{D}\bar{K}$ system is $1(1/2^+)$ considering only S-wave interactions. More specific configurations used in the present study can be referred in Table~\ref{Conf}.

\section{Prediction of a hidden charm pentaquark state with strangeness $P_{cs}^*$ and its strong decay}
\label{Results}
In this section, we predict the existence of a $\Sigma_c \bar{D}\bar{K}$ bound state and study its strong decays via triangle diagrams. The masses of the particles we used can be referred in Table~\ref{table0}.

\subsection{Prediction of a $\Sigma_{c}\bar{D}\bar{K}$ bound state as a hidden charm pentaquark state with strangeness}

In this subsection, we study the likely existence of a
$\Sigma_{c}\bar{D}\bar{K}$ three-body bound state formed with the regularized two-body potentials specified above. To estimate the uncertainties caused by the regulator, we vary the cutoff $R_c$ between  0.5 and 2.0 fm. The $\Sigma_{c}\bar{D}$ and $DK$ interactions are attractive enough to form bound states, namely $P_c(4312)$ and $D_{s0}^*(2317)$, respectively. Thus we determine the couplings $C_a$ and $C_b$ of these two interactions by reproducing the binding energies with respect to the corresponding thresholds. The results are summarized in Table~\ref{BE}, where the binding energy of the $\Sigma_{c}\bar{D}$ bound state is fixed at $8.9$ MeV and that of $DK$ is $45$ MeV. For the $\Sigma_{c}\bar{K}$ sub system, there is no direct experimental data, but one can resort to chiral perturbation theory and relate the $\Sigma_{c}\bar{K}$ interaction with the $N\bar{K}$ interaction via chiral symmetry. In this case, the $\Sigma_c\bar{K}$ interaction can form a bound state, $\Xi_{c}^*$, with a binding energy ranging from 37.7 to 71.3 MeV dependent on the cutoff $R_c$. It is interesting to note  that once the cutoff is determined via the $\Lambda(1405)$, one can find a $\Sigma_c\bar{K}$ bound state, $\Xi_{c}^*$, whose binding energy is approximately twice that of $\Lambda(1405)$ as a $N\bar{K}$ bound state.~\footnote{The $\Xi_{c}^*$ system (with quantum numbers $I=1/2$, $J=1/2$) has been studied in other works. In Ref.~\cite{Lu:2014ina}, a  state with a mass of $2695$ MeV is predicted in a coupled channel ($\Sigma_{c}\bar{K}-\Omega_cK-\Xi_c'\pi-\Xi_c'\eta$) study. In Ref.~\cite{Lutz:2003jw}, two states with the same quantum numbers are predicted with  masses 2830 and 3120 MeV respectively. In Ref.~\cite{Hofmann:2005sw}, the authors obtained 5 states with masses ranging from 2672 to 4443 MeV in a coupled-channel study of crypto-exotic baryons with charm based on chiral symmetry and large-$N_c$ QCD. }

\begin{table}[!h]
		\caption{Binding energies of the three-body $\Sigma_{c}\bar{D}\bar{K}$ system and the three two-body subsystems (in units of MeV) for three cutoffs $R_c$ (in units of fm).}
		\label{BE}
\begin{tabular}{c c c c c}
	\hline
	\hline
	$R_c$&$B_2(DK)$&$B_2(\Sigma_{c}\bar{D})$&$B_2(\Sigma_{c}\bar{K})$& $B_3(\Sigma_{c}\bar{D}\bar{K})$\\
	\hline
	0.5&45&8.9&71.3&102.8\\
	1.0&45&8.9&47.6&77.8\\
	2.0&45&8.9&37.7&67.5\\
	\hline
	\hline
\end{tabular}
\end{table}

For the $\Sigma_{c}\bar{D}\bar{K}$ system, we indeed find a bound state with quantum numbers $I(J^P)=1(1/2^+)$ and a binding energy $77.8^{+25}_{-10.3}$ MeV, see Table~\ref{BE}. The central value is obtained with $R_c=1.0$ fm while the uncertainties are obtained by taking $R_c=0.5$ and $2.0$ fm. Although the binding energy of the three-body bound state is cutoff dependent, the prediction on the existence of the $\Sigma_{c}\bar{D}\bar{K}$ bound state is robust~\footnote{Since the main uncertainties come from the $\Sigma_c\bar{K}$ interaction, we decrease its strength to only one-tenth of that of the $N\bar{K}$ interaction, and find that the $\Sigma_{c}\bar{D}\bar{K}$ system  still binds with a cutoff $R_c=1$ fm, though the two-body $\Sigma_c\bar{K}$ system already becomes unbound.}.

\begin{table}[!h]
	\caption{Root mean square radius (in units of fm) of the predicted $\Sigma_{c}\bar{D}\bar{K}$ bound states for different cutoffs $R_C$ (in units of fm).}
	\label{RMS}
	\begin{tabular}{c c c c}
		\hline
		\hline
		$R_c$&$r_3(DK)$&$r_3(\Sigma_{c}\bar{D})$&$r_3(\Sigma_{c}\bar{K})$\\
		\hline
		0.5&0.98&0.94&0.84\\
		1.0&1.29&1.28&1.27\\
		2.0&1.78&1.78&1.91\\
		\hline
		\hline
	\end{tabular}
\end{table}

In Table~\ref{RMS}, we show the root mean square (RMS) radii of the predicted $\Sigma_{c}\bar{D}\bar{K}$ bound state. The RMS radius of the $DK$ sub system in this bound state ranges from $0.98$ to 1.78 fm, increasing with the cutoff $R_c$, while those of the $\Sigma_{c}\bar{D}$ and $\Sigma_{c}\bar{K}$ sub systems ranges from 0.94 to 1.78 fm and from 0.84 to 1.91 fm, respectively.
The RMS radii of the $\Sigma_{c}\bar{D}\bar{K}$ bound state are strongly dependent on the cutoff $R_c$, because $R_c$ determines the effective interaction range.

\begin{table}[!h]
	\caption{Expectation values of the Hamiltonian (potential and kinetic energies)  (in units of MeV) of the predicted $\Sigma_{c}\bar{D}\bar{K}$ bound state for different cutoffs $R_C$ (in units of fm). The values in brackets are the specific potential weighs with respect to the total potential. }
	\label{EV1}
	\begin{tabular}{c c c c c}
		\hline
		\hline
		$R_c$&$\langle V_{DK}\rangle$&$\langle V_{\Sigma_{c}\bar{D}}\rangle$&$\langle V_{\Sigma_{c}\bar{K}}\rangle$&$\langle T\rangle$ \\
		\hline
		0.5&$-167.8(42.0\%)$&$-2.3(0.6\%)$&$-229.2(57.4\%)$&$296.5$\\
		1.0&$-104.8(51.5\%)$&$-2.1(1.0\%)$&$-96.6(47.5\%)$&$125.7$\\
		2.0&$-73.4(59.0\%)$&$-1.9(1.5\%)$&$-49.0(39.5\%)$&$56.7$\\
		\hline
		\hline
	\end{tabular}
\end{table}

In Table~\ref{EV1}, we present the expectation values of the Hamiltonian (potentials and kinetic energies) of the predicted $\Sigma_{c}\bar{D}\bar{K}$ bound state for different cutoffs, and give the weights of the  two-body potentials with respect to the total potential.
Although the absolute expectation values are strongly cutoff dependent, the relative weights of these potentials are rather stable. More specifically, the expectation value of the weight of the $DK$ potential is from $42\%$ to $59\%$, and those of the $\Sigma_{c}\bar{D}$ and $\Sigma_{c}\bar{K}$ potentials are from $0.6\%$ to $1.5\%$ and from $39.5\%$ to $57.4\%$, respectively. This indicates that to the $\Sigma_{c}\bar{D}\bar{K}$ bound state, the $DK$ and $\Sigma_{c}\bar{K}$ interaction contribute the most, which are dominant in this three-body system, while the $\Sigma_{c}\bar{D}$ interaction contributes the least, consistent with their interaction strengths.

\subsection{Two-body strong decays of $P_{cs}^{\ast}(4757)$}

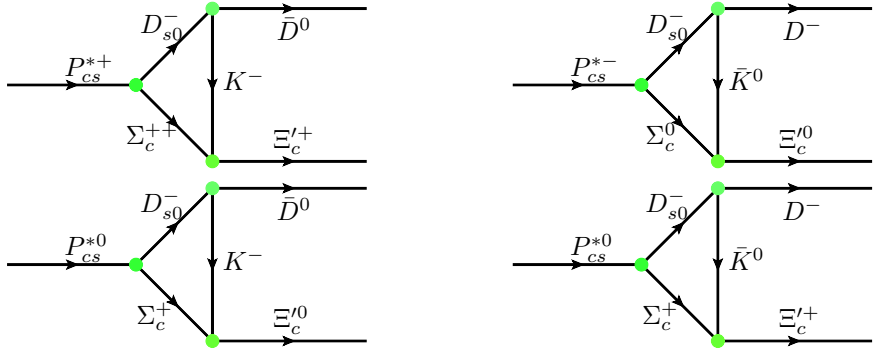
\begin{figure}[!h]
\begin{center}
\begin{tabular}{cc}
\begin{minipage}[t]{0.4\linewidth}
\begin{center}
\begin{overpic}[scale=.6]{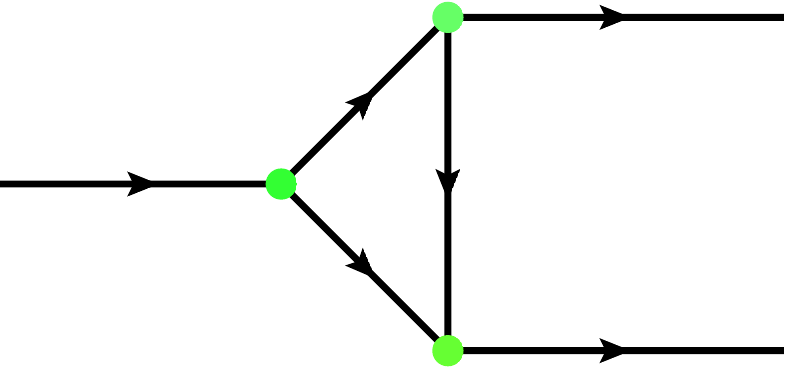}
		\put(74,6){$\Xi_c^{\prime+}$}
		
		\put(33,7){$\Sigma_c^{++}$}
		
		\put(37,38){${D}_{s0}^{-}$}
		
		\put(16,26){$P_{cs}^{\ast+}$ }
		\put(75,36){$\bar{D}^{0}$} \put(60,22){$K^{-}$}
\end{overpic}
\end{center}
\end{minipage}
&
\begin{minipage}[t]{0.4\linewidth}
\begin{center}
\begin{overpic}[scale=0.6]{triangle.pdf}
		\put(74,6){$\Xi_c^{\prime0}$}
		
		\put(37,7){$\Sigma_c^{0}$}
		
		\put(37,38){${D}_{s0}^{-}$}
		
		\put(16,26){$P_{cs}^{\ast-}$ }
		\put(75,36){${D}^{-}$} \put(60,22){$\bar{K}^{0}$}
\end{overpic}
\end{center}
\end{minipage}
\\ \nonumber
\begin{minipage}[t]{0.4\linewidth}
\begin{center}
\begin{overpic}[scale=0.6]{triangle.pdf}
		\put(74,6){$\Xi_c^{\prime0}$}
		
		\put(36,7){$\Sigma_c^{+}$}
		
		\put(37,38){${D}_{s0}^{-}$}
		
		\put(16,26){$P_{cs}^{\ast0}$ }
		\put(75,36){$\bar{D}^{0}$} \put(60,22){$K^{-}$}
\end{overpic}
\end{center}
\end{minipage}
&
\begin{minipage}[t]{0.4\linewidth}
\begin{center}
\begin{overpic}[scale=0.6]{triangle.pdf}
		\put(74,6){$\Xi_c^{\prime+}$}
		
		\put(36,7){$\Sigma_c^{+}$}
		
		\put(37,38){${D}_{s0}^{-}$}
		
		\put(16,26){$P_{cs}^{\ast0}$ }
		\put(75,36){${D}^{-}$} \put(60,22){$\bar{K}^{0}$}
\end{overpic}
\end{center}
\end{minipage}
\end{tabular}
\caption{Decay of $P_{cs}^*$ to $\bar{D}\Xi_{c}^{\prime}$  via triangle diagrams, with the hypothesis that $P_{cs}^*(4757)$  is a bound state of $\bar{D}_{s0}(2317)\Sigma_{c}$ and $\bar{D}_{s0}(2317)$ the bound state of $\bar{D}\bar{K}$.  }
\label{decay}
\end{center}
\end{figure}

\begin{figure}[!h]
\begin{center}
\begin{tabular}{cc}
\begin{minipage}[t]{0.4\linewidth}
\begin{center}
\begin{overpic}[scale=0.6]{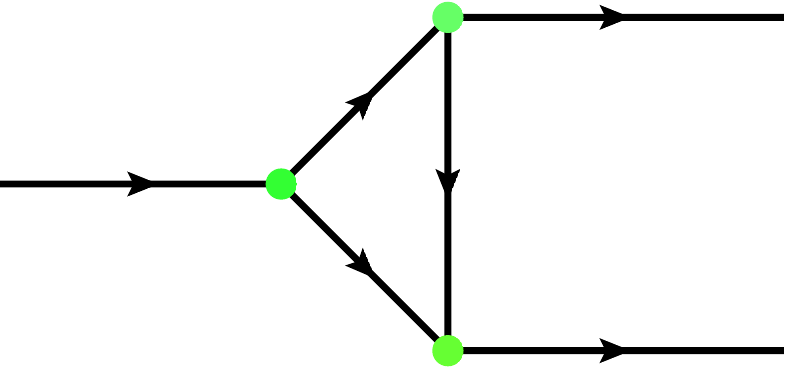}
		\put(74,6){${D}_s^{\ast-}$}
		
		\put(36,7){$\bar{D}^{0}$}
		
		\put(37,38){$\Xi_c^{\ast+}$}
		
		\put(16,26){$P_{cs}^{\ast+}$ }
		\put(75,36){$\Sigma_c^{++}$} \put(60,22){${K}^{-}$}
\end{overpic}
\end{center}
\end{minipage}
&
\begin{minipage}[t]{0.4\linewidth}
\begin{center}
\begin{overpic}[scale=0.6]{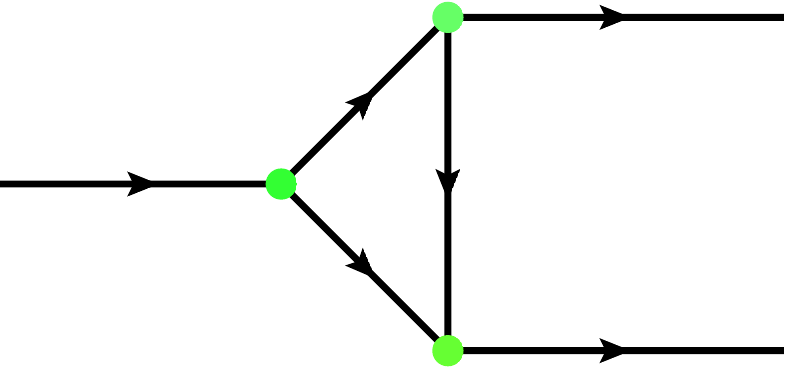}
		\put(74,6){${D}_s^{\ast-}$}
		
		\put(36,7){${D}^{-}$}
		
		\put(37,38){$\Xi_c^{\ast0}$}
		
		\put(16,26){$P_{cs}^{\ast-}$ }
		\put(75,36){$\Sigma_c^{0}$} \put(60,22){$\bar{K}^{0}$}
\end{overpic}
\end{center}
\end{minipage}
\\ \nonumber
\begin{minipage}[t]{0.4\linewidth}
\begin{center}
\begin{overpic}[scale=0.6]{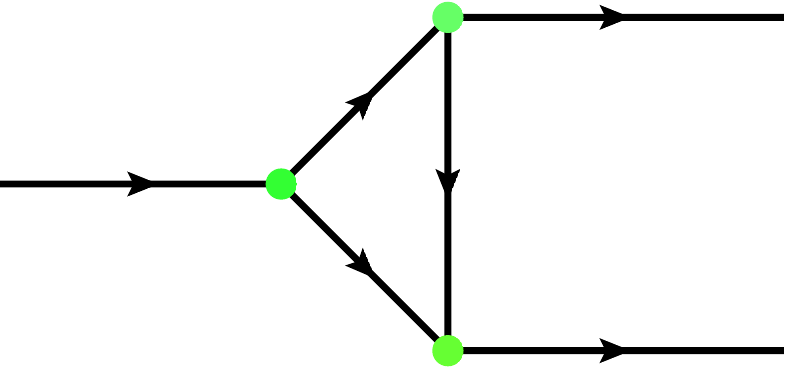}
		\put(74,6){${D}_s^{\ast-}$}
		
		\put(36,7){$\bar{D}^{0}$}
		
		\put(37,38){$\Xi_c^{\ast0}$}
		
		\put(16,26){$P_{cs}^{\ast0}$ }
		\put(75,36){$\Sigma_c^{+}$} \put(60,22){${K}^{-}$}
\end{overpic}
\end{center}
\end{minipage}
&
\begin{minipage}[t]{0.4\linewidth}
\begin{center}
\begin{overpic}[scale=0.6]{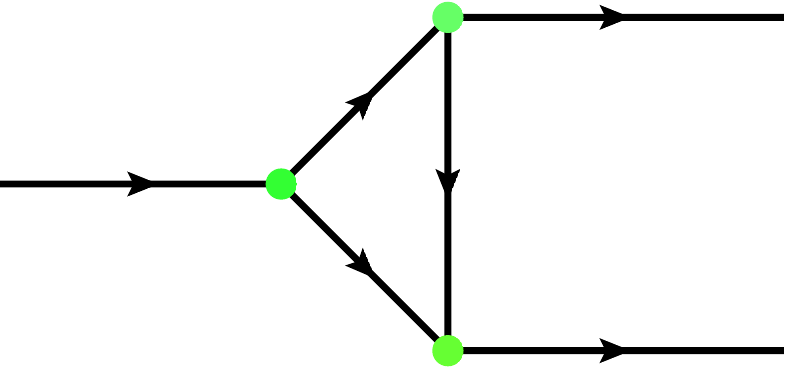}
		\put(74,6){${D}_s^{\ast-}$}
		
		\put(36,7){${D}^{-}$}
		
		\put(37,38){$\Xi_c^{\ast+}$}
		
		\put(16,26){$P_{cs}^{\ast0}$ }
		\put(75,36){$\Sigma_c^{+}$} \put(60,22){$\bar{K}^{0}$}
\end{overpic}
\end{center}
\end{minipage}
\end{tabular}
\caption{Decay of $P_{cs}^*$ to $\bar{D}_{s}^{\ast}\Sigma_{c}$  via triangle diagrams, with the hypothesis that $P_{cs}^*(4757)$  is a bound state of $\bar{D}\Xi_{c}^{\ast}$ and $\Xi_{c}^{\ast}$ the bound state of $\Sigma_{c}\bar{K}$.  }
\label{decay1}
\end{center}
\end{figure}

\begin{figure}[!h]
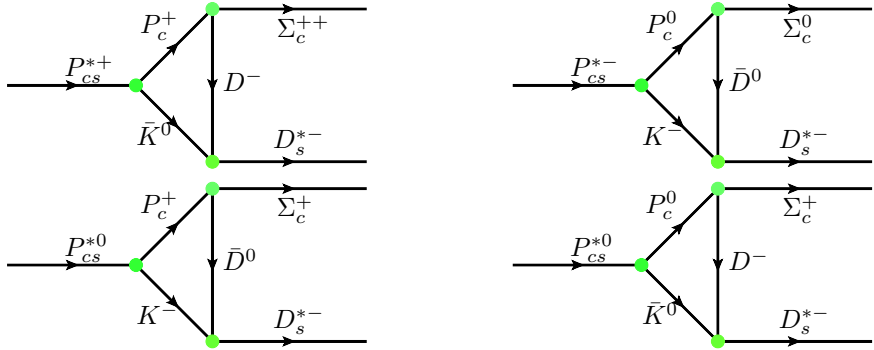

\begin{center}
\begin{tabular}{cc}
\begin{minipage}[t]{0.4\linewidth}
\begin{center}
\begin{overpic}[scale=.6]{triangle.pdf}
		\put(74,6){${D}_s^{\ast-}$}
		
		\put(36,7){$\bar{K}^{0}$}
		
		\put(37,38){$P_c^{+}$}
		
		\put(16,26){$P_{cs}^{\ast+}$ }
		\put(75,37){$\Sigma_c^{++}$} \put(60,22){${D}^{-}$}
\end{overpic}
\end{center}
\end{minipage}
&
\begin{minipage}[t]{0.4\linewidth}
\begin{center}
\begin{overpic}[scale=0.6]{triangle.pdf}
		\put(74,6){${D}_s^{\ast-}$}
		
		\put(36,7){${K}^{-}$}
		
		\put(37,38){$P_c^{0}$}
		
		\put(16,26){$P_{cs}^{\ast-}$ }
		\put(75,37){$\Sigma_c^{0}$} \put(60,22){$\bar{D}^{0}$}
\end{overpic}
\end{center}
\end{minipage}
\\ \nonumber
\begin{minipage}[t]{0.4\linewidth}
\begin{center}
\begin{overpic}[scale=0.6]{triangle.pdf}
		\put(74,6){${D}_s^{\ast-}$}
		
		\put(36,7){${K}^{-}$}
		
		\put(37,38){$P_c^{+}$}
		
		\put(16,26){$P_{cs}^{\ast0}$ }
		\put(75,37){$\Sigma_c^{+}$} \put(60,22){$\bar{D}^{0}$}
\end{overpic}
\end{center}
\end{minipage}
&
\begin{minipage}[t]{0.4\linewidth}
\begin{center}
\begin{overpic}[scale=0.6]{triangle.pdf}
		\put(74,6){${D}_s^{\ast-}$}
		
		\put(36,7){$\bar{K}^{0}$}
		
		\put(37,38){$P_c^{0}$}
		
		\put(16,26){$P_{cs}^{\ast0}$ }
		\put(75,37){$\Sigma_c^{+}$} \put(60,22){${D}^{-}$}
\end{overpic}
\end{center}
\end{minipage}
\end{tabular}
\caption{Decay of $P_{cs}^*$ to $\bar{D}_{s}^{\ast}\Sigma_{c}$  via triangle diagrams, with the hypothesis that $P_{cs}^*(4757)$  is a bound state of $P_{c}(4312)\bar{K}$ and $P_{c}(4312)$ the bound state of $\Sigma_{c}\bar{D}$.  }
\label{decay2}
\end{center}
\end{figure}

In the following we explore the strong decays of  $P_{cs}^{\ast}$ via triangle diagrams. From our above study, it is clear that   $P_{cs}^{\ast}$  can not only be viewed as a $\Sigma_c\bar{D}\bar{K}$ bound state, but also  be regarded as three kinds of quasi two-body bound states, $\bar{D}_{s0}(2317)\Sigma_{c}$, $\bar{D}\Xi_{c}^{\ast}$, and $P_{c}(4312)\bar{K}$. Therefore, the decay of $P_{cs}^{\ast}$ can also proceed through three modes as shown in Figs.~\ref{decay}-\ref{decay2}. Assuming that  $P_{cs}^{\ast}$ is mainly made of $\bar{D}_{s0}(2317)\Sigma_{c}$ and $\bar{D}_{s0}(2317)$ is a bound state of $\bar{D}\bar{K}$, $P_{cs}^{\ast}$ can decay to $\bar{D}\Xi_{c}^{\prime}$ via the triangle diagrams  shown in Fig.~\ref{decay}. Using the same mechanism we display the other two decay modes of $P_{cs}^{\ast}$  as shown in Fig.~\ref{decay1} and Fig.~\ref{decay2}.
To give a quantitative estimate of the decays of $P_{cs}^{\ast}$ in these processes, we employ the effective Lagrangian approach to calculate their decay widths, which has been widely used to explore strong decays of hadronic molecules, see, e.g., Refs.~\cite{Huang:2019qmw,Lu:2016nnt,Lin:2019qiv,Xiao:2019mvs}.  The relevant Lagrangians are
\begin{eqnarray}
\mathcal{L}_{P_{cs}^{\ast}\bar{D}_{s0}(2317)\Sigma_{c}}&=&-i g_{P_{cs}^{\ast}\bar{D}_{s0}(2317)\Sigma_{c}}P_{cs}^{\ast}\bar{D}_{s0}(2317)\Sigma_{c}, \\ \nonumber
\mathcal{L}_{P_{cs}^{\ast}P_{c}(4312)\bar{K}}&=& g_{P_{cs}^{\ast}P_{c}(4312)\bar{K}}P_{cs}^{\ast}P_{c}(4312)\bar{K},\\ \nonumber
\mathcal{L}_{P_{cs}^{\ast}\Xi_{c}^{\ast}\bar{D}}&=& g_{P_{cs}^{\ast}\Xi_{c}^{\ast}\bar{D}}P_{cs}^{\ast}\Xi_{c}^{\ast}\bar{D},\\ \nonumber
\mathcal{L}_{\bar{D}_{s0}(2317)\bar{D}\bar{K}}&=& g_{\bar{D}_{s0}(2317)\bar{D}\bar{K}}\bar{D}_{s0}(2317)\bar{D}\bar{K}, \\ \nonumber
\mathcal{L}_{P_{c}(4312)\bar{D}\Sigma_{c}}&=&-ig_{P_{c}(4312)\bar{D}\Sigma_{c}}P_{c}(4312)\bar{D}\Sigma_{c}, \\ \nonumber
\mathcal{L}_{\Xi_{c}^{\ast}\bar{K}\Sigma_{c}}&=&-ig_{\Xi_{c}^{\ast}\bar{K}\Sigma_{c}}\Xi_{c}^{\ast}\bar{K}\Sigma_{c}, \\ \nonumber
\mathcal{L}_{\Sigma_{c}\Xi_{c}^{\prime}\bar{K}}&=i& g_{\Sigma_{c}\Xi_{c}^{\prime}\bar{K}}\bar{\Sigma}_{c}\gamma_{\mu}\gamma^{5}\partial^{\mu}K \Xi_{c}^{\prime}, \\ \nonumber
\mathcal{L}_{\bar{D}\bar{D}_{s}^{\ast}\bar{K}}&=&ig_{\bar{D}\bar{D}_{s}^{\ast}\bar{K}}D_{s}^{\ast\mu}(\bar{D}\partial_{\mu}\bar{K}-\bar{K}\partial_{\mu}\bar{D}),
\end{eqnarray}
    where the couplings of each vertex are classified into two scenarios, molecular type that a particle is assumed as a bound state of the other two particles and scattering type that a particle can change into another particle by exchanging  a light meson. For the molecular type the couplings can be estimated by the Weinberg compositeness condition~\cite{Faessler:2007us,Faessler:2007gv}, where the renormalization constant of the composite particle should be zero. To remove the ultraviolet divergence of the loop diagrams, we choose a Gaussian form factor $\exp(-p_{E}^2/\Lambda^2)$, where $P_{E}$ is the Euclidean Jacobi momentum and $\Lambda$ characterizes the distribution of the molecular components inside the molecule. The cutoff value is often chosen to be $\Lambda=$1 GeV~\cite{Huang:2019qmw,Xiao:2019mvs,Wu:2020job}.   With this value the coupling between $P_{cs}^{\ast}$ and its component $\bar{D}_{s0}(2317)$ and $\Sigma_{c}$ is found to be $g_{P_{cs}^{\ast}\bar{D}_{s0}(2317)\Sigma_{c}}=3.65$. Since $\bar{D}_{s0}(2317)$ is treated as a $\bar{D}\bar{K}$ bound state in this work, the corresponding coupling can be determined as $g_{\bar{D}_{s0}(2317)\bar{D}\bar{K}}=7.35$ GeV, whose value is a little bit smaller than  that obtained in other approaches~\cite{Gamermann:2006nm,Guo:2006fu}.  The other relevant couplings  can also be determined in the same way as $g_{P_{cs}^{\ast}P_{c}(4312)\bar{K}}=4.11$, $g_{P_{cs}^{\ast}\bar{D}\Xi_{c}^{\ast}}=3.19$,  $g_{P_{c}(4312)\bar{D}\Sigma_{c}}=2.24$, and $g_{\Xi_{c}^{\ast}\bar{K}\Sigma_{c}}=3.74$. For the scattering type vertices the couplings of $\Sigma_{c}\Xi_{c}^{\prime}\bar{K}$  and $\bar{D}D^{\ast}\bar{K}$ can be determined as $g_{\Sigma_{c}\Xi_{c}^{\prime}\bar{K}}=9.01$ and $g_{\bar{D}\bar{D}_{s}^{\ast}\bar{K}}=4.54$ from the couplings of $\Sigma_{c}\Sigma_{c}\pi$ and $\bar{D}\bar{D}^{\ast}\pi$,  via $SU(3)$-flavor symmetry.

With the above vertices determined, we obtain the amplitudes of the corresponding triangle diagrams as
\begin{eqnarray}
\label{amp}
i\mathcal{M}&=&g_{P_{cs}^{\ast}\bar{D}_{s0}(2317)\Sigma_{c}}g_{\bar{D}_{s0}(2317)\bar{D}\bar{K}}g_{\Sigma_{c}\Xi_{c}^{\prime}\bar{K}}\int\frac{d^{4}q}{(2\pi)^{4}}\bar{u}_{\Xi_{c}^{\prime}}\gamma^{\mu}q_{\mu}\gamma_{5}\frac{/\!\!\!k_{1}+m_{\Sigma_{c}}}{k_{1}^{2}-m_{\Sigma_{c}}^2}\frac{1}{k_{2}^{2}-m_{\bar{D}_{s0}^*}^{2}}\frac{1}{q^{2}-m_{\bar{K}}^{2}}u_{P_{cs}^{\ast}}, \\ \nonumber
i\mathcal{M}&=&g_{P_{cs}^{\ast}\Xi_{c}^{\ast}\bar{D}}g_{\Xi_{c}^{\ast}\bar{K}\Sigma_{c}}g_{\bar{D}\bar{K}\bar{D}_{s}^{\ast}}\int\frac{d^{4}q}{(2\pi)^{4}}\bar{u}_{\Sigma_{c}}\frac{/\!\!\!k_{1}+m_{\Xi_{c}^{\ast}}}{k_{1}^{2}-m_{\Xi_{c}^{\ast}}^2}\frac{2\varepsilon_{\bar{D}_{s}^{\ast}}\cdot q}{k_{2}^{2}-m_{\bar{D}}^{2}}\frac{1}{q^{2}-m_{\bar{K}}^{2}}u_{P_{cs}^{\ast}},
\\ \nonumber
i\mathcal{M}&=&g_{P_{cs}^{\ast}P_{c}\bar{K}}g_{P_{c}\bar{D}\Sigma_{c}}g_{\bar{D}\bar{K}\bar{D}_{s}^{\ast}}\int\frac{d^{4}q}{(2\pi)^{4}}\bar{u}_{\Sigma_{c}}\frac{/\!\!\!k_{1}+m_{P_{c}}}{k_{1}^{2}-m_{P_{c}}^2}\frac{2\varepsilon_{\bar{D}_{s}^{\ast}}\cdot q}{k_{2}^{2}-m_{\bar{K}}^{2}}\frac{1}{q^{2}-m_{\bar{D}}^{2}}u_{P_{cs}^{\ast}},
\end{eqnarray}
where $k_{1}$, $k_{2}$ and $q$ denote the momenta of particles appearing in the triangle diagrams, $u_{P_{cs}^{\ast}}$ and $\bar{u}_{\Sigma_{c}}$ represent the initial and final spinors, respectively, and $\varepsilon_{\bar{D}_{s}^{\ast}}$ is the polarization vector of $\bar{D}_{s}^{\ast}$.   In addition, to eliminate the ultraviolet divergence, we also add the Gaussian form factor in the above amplitudes.  The partial decay width of $P_{cs}^{\ast}$ can be finally obtained by
\begin{eqnarray}
\Gamma=\frac{1}{2J+1}\frac{1}{8\pi}\frac{|\vec{p}|}{m_{P_{cs}^{\ast}}^2}\bar{|\mathcal{M}|}^{2},
\end{eqnarray}
where $J$ is the total angular momentum of the initial state $P_{cs}^{\ast}$, the overline indicates the sum over the polarization vectors of final states, and $|\vec{p}|$ is the momentum of either final state in the rest frame of  $P_{cs}^{\ast}$.

\begin{table}[!h]
	\caption{Partial decay widths (in units of MeV) of the predicted $P_{cs}^{\ast}$ as  three quasi two-body bound states $ \bar{D}_{s0}\Sigma_{c}$,  $ \bar{D}\Xi_{c}^{\ast}$, and $   \bar{K}P_{c}$ with $\Lambda=0.8\sim 1.2$ GeV. }
	\label{Decay}
	\begin{tabular}{c c c c c}
		\hline
		\hline
		Modes &$\bar{D}_{s0}\Sigma_{c} \to \bar{D}\Xi_{c}^{\prime}$&  $ \bar{D}\Xi_{c}^{\ast}\to \bar{D}_{s}^{\ast}\Sigma_{c}$&$  \bar{K}P_{c} \to  \bar{D}_{s}^{\ast}\Sigma_{c}$ \\
		\hline
$P_{cs}^{\ast+}$& $92.0\sim 99.5$  &  $25.4\sim 28.0$ &    $0.2\sim 0.3$ \\
$P_{cs}^{\ast-}$& $92.0\sim 99.5$ &$25.4\sim 28.0$&$0.2\sim 0.3$ \\
$P_{cs}^{\ast0}$&$46.0\sim 49.7$&    $25.4\sim 28.0$&$0.2\sim 0.3$ \\
		\hline
		\hline
	\end{tabular}
\end{table}

 Within the molecular picture studied in the present work, $P_{cs}^{\ast}$ can decay via three possible  modes, the results are presented in Table~\ref{Decay}.
If  $P_{cs}^{\ast}$ decays via $\bar{D}_{s0}(2317)\Sigma_{c}$ as shown in Fig.~\ref{decay},  we find a partial decay width of $P_{cs}^{\ast+}\to\bar{D}^{0}\Xi_{c}^{\prime0}$ 96.3 MeV with $\Lambda=1$ GeV.  The widths of its isospin partner $P_{cs}^{\ast-}\to{D}^{-}\Xi_{c}^{\prime0}$ and $P_{cs}^{\ast0}\to{D}^{-}\Xi_{c}^{\prime+}/\bar{D}^{0}\Xi_{c}^{\prime0}$ are 96.3 MeV and 48.2 MeV, respectively. If it decays via the second mode  shown in Fig.~\ref{decay1}, the decay widths of  $P_{cs}^{\ast-(+)}\to\Sigma_{c}^{0(++)}D_{s}^{\ast-}$ and $P_{cs}^{\ast0}\to\Sigma_{c}^{+}D_{s}^{\ast-}$ are both  25.7 MeV.
In the third mechanism  shown in Fig.~\ref{decay2} the final states of the $P_{cs}^{\ast}$ decay are the same as those of the second mechanism, while the decay widths are much smaller, less than 1 MeV.  Clearly the dominate decay mode is the first mechanism shown in Fig.~\ref{decay}. The final states are experimentally accessible in the $\bar{D}^{0}\Xi_{c}^{\prime+}$ or ${D}^{-}\Xi_{c}^{\prime0}$ channel.

\section{Summary and outlook}
\label{Summary}

In this work, we employed the Gaussian Expansion Method to study the three-body $\Sigma_{c}\bar{D}\bar{K}$ system.
For the $\Sigma_{c}\bar{D}$ interaction, we refereed to the contact-range effective field theory, and renormalized the potential with a Gaussian regulator and a cutoff $R_c$ in coordinate space. For the $DK$ interaction, the most important contribution is the Weinberg-Tomozawa term, which is also a contact-range potential. We made the same renormalization procedure following the case of $\Sigma_{c}\bar{D}$. We chose the cutoff ranging from 0.5 to 2.0 fm and determined the couplings of $\Sigma_{c}\bar{D}$ and $DK$ potentials by reproducing $P_c(4312)$ and $D_{s0}^*(2317)$, respectively.
The $\Sigma_{c}\bar{K}$ interaction is related to the $N\bar{K}$ one via chiral symmetry, and is also found to generate a bound state with a binding energy of $38-71$ MeV depending on the chosen cutoff.

With the regularized two-body interactions, the three-body $\Sigma_{c}\bar{D}\bar{K}$ system is found to form a bound state, denoted as $P_{cs}^*(4739)$, with quantum numbers $I(J^P)=1(1/2^+)$ and a binding energy $77.8^{+25}_{-10.3}$ MeV. The RMS radii and Hamiltonian expectation values of the predicted bound were also presented, which showed that the $DK$ and $\Sigma_c\bar{K}$ interactions contribute most to the formation of the three-body bound state.

Based on the molecular nature of the predicted three-body bound state, we studied the two-body open-charm decays of $P_{cs}^*(4739)$ with the effective Lagrangian method via triangle diagrams. We found that the $P_{cs}^*(4739)$ state can decay into $\Xi_{c}^{'}\bar{D}$ and $\bar{D}_s^*\Sigma_c$ with  partial decay widths of a few tens' of MeV.

It is expected that the isovector pentaquark $P_{cs}^*(4739)$ can also decay into the hidden-charm channel $J/\psi\Sigma$. Therefore it may be possible to search for the $P_{cs}^*(4739)$ in the Cabibbo-favored $\Lambda_b\to\pi J/\psi \Sigma$ decay. The corresponding search may be performed by the LHCb Collaboration. However, it is not easy to identify the $\Sigma$ baryon at hadron-hadron colliders and thus this poses a challenge to experiments.

The current study can be easily extended to the $\bar{\Sigma}_cDK$ system with a minus $C$ parity. Employing the heavy quark symmetry, the study can also be extended to the $\Sigma_{c}\bar{D}^*\bar{K}$ system.
These predictions of the $\Sigma_{c}\bar{D}^{(*)}\bar{K}$ ($\bar{\Sigma}_cD^{(*)}K$) bound state provide  the first  hidden charm fermionic three-body molecules,
which are likely to be found at
the current facilities, especially considering the successful discoveries of the $P_c$ and $P_{cs}$ states by the LHCb Collaboration. Thus we encourage our experimental colleagues to search for them.

\bibliography{SigmacDbarKbar}

\end{document}